%% file: main.tex
\documentclass{article}
\usepackage{ijcai21}

\usepackage{times}

\usepackage{soul}
\usepackage{url}
\usepackage[hidelinks]{hyperref}
\usepackage[utf8]{inputenc}
\usepackage[small]{caption}
\usepackage{graphicx}
\usepackage{amsmath}
\usepackage{booktabs}
\usepackage{dsfont}
\usepackage{amsfonts}
\usepackage{tikz}
\usepackage{pgfplots}

\usepackage[left=37pt,right=37pt,top=54pt,bottom=113pt]{geometry}

\xdefinecolor{tucol1}{rgb}{1.0 0.5 0.05}
\xdefinecolor{tucol2}{rgb}{0.52,0.72,0.10}
\xdefinecolor{tucol3}{rgb}{0.0 0.0 0.8}
\xdefinecolor{tucol4}{rgb}{0.8 0.8 0}
\xdefinecolor{tucolv}{rgb}{0.1 0.35 0.17}
\xdefinecolor{tucol5}{rgb}{0.8 0 0.8}
\xdefinecolor{tucol6}{rgb}{0 0.8 0.8}
\xdefinecolor{tucol7}{rgb}{0.7 0.8 0.1}
\xdefinecolor{tucol8}{rgb}{0.2 0.7 0.34}
\xdefinecolor{tucol9}{rgb}{0.7 0.8 0.3}
\xdefinecolor{softgreen}{rgb}{0.49,0.50,0.47}

\DeclareMathOperator{\given}{|}

\title{Human Activity Recognition using Attribute-Based Neural Networks and Context Information}

\author{
Stefan L\"udtke$^1$\footnote{These authors contributed equally to the paper.}\and
Fernando Moya Rueda$^{2*}$\and
Waqas Ahmed$^{1}$\and
Gernot A. Fink$^2$\And
Thomas Kirste$^1$
\affiliations
$^1$Institute of Visual \& Analytic Computing, University of Rostock, Germany\\
$^2$Department of Computer Science, TU Dortmund University, Germany\\
\emails
\{stefan.luedtke2, waqas.ahmed, thomas.kirste\}@uni-rostock.de,\\
\{fernando.moya, gernot.fink\}@tu-dortmund.de
}

\begin{document}

\maketitle

\begin{abstract}
We consider human activity recognition (HAR) from wearable sensor data in manual-work processes, like warehouse order-picking. Such structured domains can often be partitioned into distinct \emph{process steps}, e.g., packaging or transporting. Each process step can have a different prior distribution over \emph{activity classes}, e.g., standing or walking, and different system dynamics. 

Here, we show how such context information can be integrated systematically into a deep neural network-based HAR system. Specifically, we propose a hybrid architecture that combines a deep neural network---that estimates high-level movement descriptors, \emph{attributes}, from the raw-sensor data---and a shallow classifier, which predicts activity classes from the estimated attributes and (optional) context information, like the currently executed process step. 

We empirically show that our proposed architecture increases HAR performance, compared to state-of-the-art methods. Additionally, we show that HAR performance can be further increased when information about process steps is incorporated, even when that information is only partially correct. 




\end{abstract}

\section{Introduction}

\begin{figure}[t]
        \centering
        \includegraphics[width=\columnwidth]{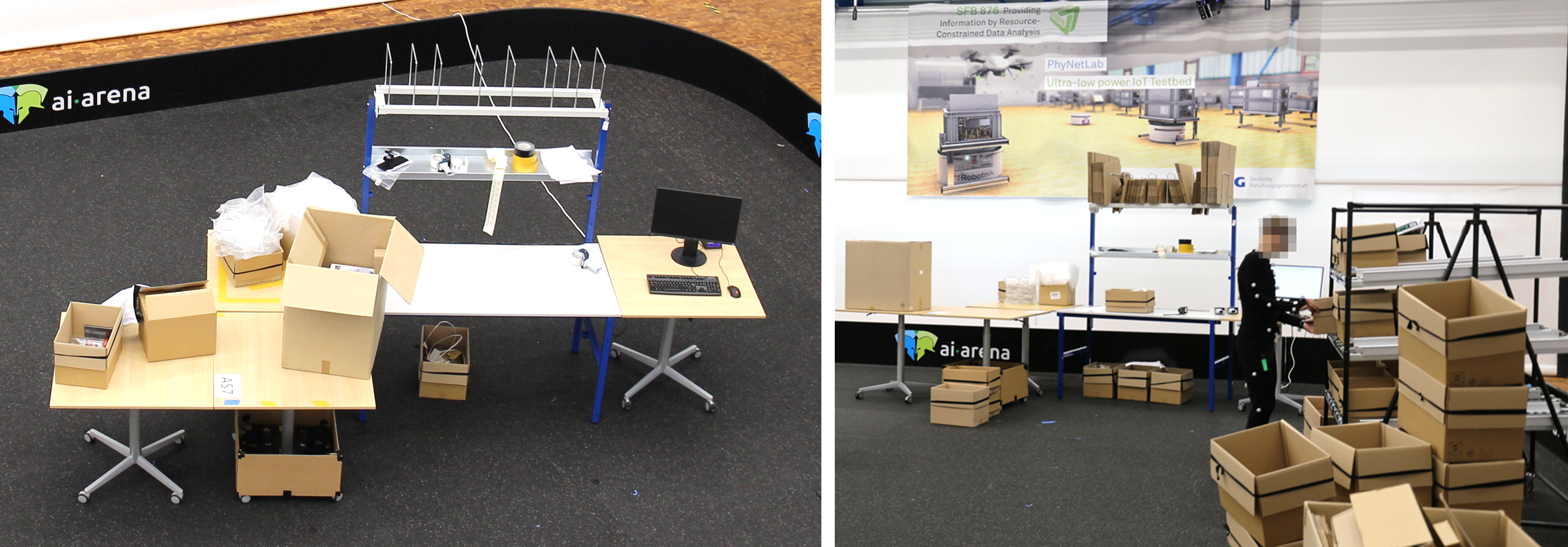}
        \caption{Physical laboratory set-up of a warehouse packaging work station.}
        \label{fig:scenario}
    \end{figure}

    The accurate recognition of human activities from sensor data is an important task for many applications, like healthcare, sports, or for developing situation-aware assistive technologies. 
    As another example, activity recognition is relevant for the analysis and optimization of manual-work processes, like packaging in a warehouse, see Fig.~\ref{fig:scenario}, the main motivation for our work \cite{reining2018_TFSAAHOPAUMC,niemann_lara_2020}. 
    
    Human Activity Recognition (HAR) from movement data is challenging due to the intra- and inter-subject variability of human movement. 
    Recently, deep neural networks have been used successfully for multichannel time-series HAR. These networks combine the feature extraction and classification in an end-to-end approach \cite{ordonez_deep_2016,zeng_2014_CNNHARUMS,ronao_deep_2015,hammerla_deep_2016,yao_efficient_2018}. They learn non-linear and temporal relations of basic, complex, and highly dynamic human movements directly from raw-inertial sensor data. These transformations are more discriminative with respect to human action classes than hand-crafted features \cite{zeng_2014_CNNHARUMS,hammerla_deep_2016}.
    
    However, such deep neural networks are usually purely data-driven---they cannot directly make use of prior knowledge that is often available in highly structured application domains.
    For example, manual work processes, like order picking or packaging, are structured into distinct high-level \emph{process steps}, as shown in Fig.~\ref{fig:bpm}.
    Each of these steps has a unique prior distribution over activity classes and unique system dynamics. For example, a person is more likely to perform a \emph{handling} activity class when retrieving articles in the warehouse (step 3 in Fig.~\ref{fig:bpm}) than when moving the cart (step 2 in Fig.~\ref{fig:bpm}). 
    Furthermore, information about the currently executed process step is often available directly from external sources. For example, in a packaging scenario, each article and empty box are scanned before being packed, according to a shipment list. This scanning process, which is recognized in the warehouse management system, indicates a transition between process steps. 
    
    In this paper, we show how such additional \emph{context information} can be integrated systematically into a deep neural network-based HAR system. 
    Specifically, we propose a hybrid architecture that allows integrating context information without re-training the neural network from scratch: We use a deep neural network that extracts high-level movement descriptors (\emph{attributes}, like posture or usage of left or right hand) from raw wearable-sensor data. On top, a shallow classifier estimates activity classes from the attributes and optionally from the context information---like the currently executed process step, in our case. 
    This way, we can use a deep neural network independently of available context information and domain specifics. Instead, we only need to adapt the final classifier to different domains, i.e., different process models or changes in the available context information. 
    
    We evaluate this architecture on a HAR task in intra-logistics, and empirically show that it achieves increased HAR performance, compared to the state of the art, even when no context information is available. Additionally, the HAR performance increases further when context information in the form of the currently executed process step is available---even when it is just partially correct.

\section{Related Work}

    Methods of statistical pattern recognition are common for analyzing human movements from measurements from on-body devices. A standard pipeline involves pre-processing, segmentation, hand-crafted feature extraction, and classification. Recently, deep learning methods become the standard method for solving HAR problems in gesture recognition and activities of daily living \cite{ordonez_deep_2016,grzeszick2017_DNNBHAROPP,hammerla_deep_2016,ronao_deep_2015,yao_efficient_2018}.
    Deep architectures combine the feature extraction and classification in an end-to-end approach. These architectures learn the non-linear and temporal relations of basic, complex, and highly dynamic human movements. They learn non-linear transformations directly from raw-inertial data. Compared to the hand-crafted ones, these transformations are more discriminative to human actions and invariant to distortions and temporal translations \cite{hammerla_deep_2016}. The authors in \cite{ronao_deep_2015} introduced temporal convolutional neural networks (tCNNs), which carry out convolution and pooling operations along the time axis. These architectures share rather small convolution filters among all the sensors, as local temporal measurements are correlated independently of the sensor type; this is also valid assuming a normalization per sensor. An architecture that combines temporal convolutions and recurrent networks is introduced in \cite{ordonez_deep_2016}. Specifically, the authors used Long Short-Term Memory units (LSTMs), recurrent units with memory cells, and a gating system \cite{hochreiter_long_1997}. The authors in \cite{hammerla_deep_2016} utilized shallow recurrent-networks: a three-layered LSTM and a one-layered bidirectional LSTM. The authors in \cite{grzeszick2017_DNNBHAROPP} created a tCNN that adapts to the wearables per human limb. This architecture contains parallel-convolutional branches per human limb. Each convolutional branch creates a deep representation of the measurements per human limb. The network deploys a late fusion for creating a final deep representation. The authors in \cite{qian_2019} proposed an end-to-end architecture that extract temporal, body-relations and statistical features.
     
    Nonetheless, the performance of these deep learning methods did not show a significant increase compared to other areas such as image and video classification or segmentation \cite{hammerla_deep_2016,grzeszick2017_DNNBHAROPP}. HAR remains a challenging task due to large intra- and inter-class variability of human movements, i.e., humans carry out similar tasks differently. In addition, there is a broad range of human activities or movements, and there is not a standard definition nor structure for formulating a clear problem of HAR \cite{bulling_tutorial_2014}. Likewise, datasets for HAR suffer from the class-unbalance problem, where the number of samples per action class differs strongly \cite{ordonez_deep_2016}.

    Attribute representations help solve zero-shot learning and transfer learning. High-level attribute representations are semantic descriptions that describe categories, e.g., in object and scene recognition problems \cite{Cheng_TZLHARUSASM,lampert_attribute-based_2014,zheng_submodular_2017} and words in document analysis. In object recognition, attributes can be the shape, color, texture, size of objects, or even geographic information. Different approaches have introduced semantic descriptions of activities as representations for solving HAR \cite{Cheng_TZLHARUSASM,arif_ul_alam_unseen_2017,zheng_submodular_2017}. In \cite{Cheng_TZLHARUSASM}, the authors proposed to use semantic attributes for recognizing unseen activities. In \cite{zheng_submodular_2017}, human-annotated attributes and data-driven attributes are combined for solving HAR in sports videos. They selected a subset from both attribute groups maximizing the discrimination capability of attributes for distinguishing different sets of classes. The authors in \cite{arif_ul_alam_unseen_2017} designed a hierarchical representation of human-activity taxonomy based on semantic descriptions in the context of smart-home applications.
    
    Precisely, attribute-based representations have been deeply explored on HAR in the manual order picking process in \cite{reining2018_TFSAAHOPAUMC}. Attribute representations are beneficial for dealing with the versatility of activities. The authors in \cite{reining2018_TFSAAHOPAUMC} compared the performance of deep architectures trained using different attribute representations, evaluating their quantitative performance and quality from the perspective of practical application. The authors in \cite{reining2018_ARHARMOPA} tested different attribute representations, expert-given and random, for solving zero-shot learning in HAR. Unseen activities were described using attributes that are shared with the seen activities. Expert-given attribute representations performed better than a random one. The latter is created following the conclusions in \cite{moya2018_LARHAR}. A semantic relation between attributes and activities enhances HAR not only quantitatively with regards to performance but also guarantees a transfer of the attributes between activities by domain experts. In this preliminary work, the mapping between activity classes and attribute representations was one-to-one. Fig.~\ref{fig:attribute_based_representation} presents an example of attribute-based representation for HAR.

    Deep neural networks do not consider domain knowledge about the causal relation of activities that is often available in highly structured domains. To overcome this limitation, a combination of symbolic reasoning methods (Computational State-Space Models, CSSM \cite{kruger2014computational}) and deep neural networks is introduced in \cite{fmoya_CSRDLHAR}. The CSSM models prior knowledge about the high-level, causal structure of the domain, and the deep neural network acts as observation model, relating the sensor data to CSSM states. 

\section{Methods}

    \input{figures/attribute_based_representation}


    We propose a hybrid HAR method, consisting of a deep neural network that predicts movement descriptors (attributes) from the sensor data, and a model that predicts activity classes from the attribute estimates. 
    
    In the following, we first present the architecture of the neural network in more detail. Afterwards, we discuss different options for representing and learning the relationship between attributes and activity classes. 
    Finally, we show how context information can be integrated into the classification system.


    %

    \subsection{Attribute-based Deep Neural Network for HAR}
    \label{subsec:network}
    
        We deploy the temporal convolutional neural network (tCNN) from \cite{yang2015_DCNNMTSHAR,grzeszick2017_DNNBHAROPP}. It has shown to perform relatively well despite its simplicity \cite{grzeszick2017_DNNBHAROPP,moya2018_LARHAR}. A tCNN is an end-to-end architecture composed of feature extractors and a classifier, either a softmax or a sigmoid. The architecture processes sequences of size $[T,W]$, with $T$ the sequence length and $W$ the number of sequence channels. The tCNN contains four convolutional layers, no downsampling, and two fully connected layers, and a classifier. The convolutional layers are composed of $64$ filters of size $[5\times 1]$, performing convolutions along the time axis. The first and second fully connected layers contain $128$ units. Depending on the task, the tCNN will have a softmax layer for activity classification or a sigmoid layer for attribute classification \cite{moya2018_LARHAR}. Here, a sigmoid activation function replaces the usual softmax layer, as we use the tCNN$_{attribute}$ to compute an attribute representation from an input sequence, rather than directly classifying it, following \cite{moya2018_LARHAR}. Using attribute representations have shown to be beneficial for HAR, following the conclusions in \cite{moya2018_LARHAR,niemann_lara_2020}. The tCNN$_{attribute}$ is trained using the binary-cross entropy loss. Fig.~\ref{fig:attribute_based_representation} presents the relation between activities, attributes, and sequence input, and Fig.~\ref{fig:method} shows the tCNN$_{attribute}$ architecture.
        
        Due to the final sigmoid layer, the tCNN$_{attribute}$ output can be interpreted as the probabilities for each attribute being present or not present in the input segment.
        More formally, the tCNN$_{attribute}$ is a function $\phi: D \rightarrow \Pi$, where $d \in D$ is a data sample, i.e., segment, and $\pi \in \Pi$ is a parameter vector for the posterior distribution over attributes $p_\phi(a \given d)$. Specifically, the probability of a binary attribute vector $a = (a_1,\dots,a_K)$ is given by a product of Bernoulli distributions:
        \begin{align}
            \label{eq:bernoulli}
            p_\phi(a \given d)  = \prod_{k=1}^K p(a_k \given d)= \prod_{k=1}^K \pi_k^{a_k} \, (1-\pi_k)^{1-a_k} 
        \end{align}

    \input{figures/method_fig}
        
    \subsection{Human Activity Recognition from Semantic Attributes}
    \label{subsec:attributes-to-class}
    
        From an application viewpoint, we are interested in the activity class $c$ instead of the attributes themselves. 
        In general, the maximum-likelihood estimate $\hat{c}$ of the activity class is given by
        \begin{equation}
        \hat{c} = \text{argmax}_c \, p(c \given d)
        \label{eq:class-estimate}
        \end{equation}
        We assume that the activity class $c$ and the sensor data $d$ are conditionally independent, given the attribute vector $a$, so that the activity class posterior can be written as
        \begin{align}
         p(c \given d) = \sum_{a \in A} p(c \given a)\, p_\phi(a \given d).
          \label{eq:attribute-class-model}
        \end{align}
        Here, $p_\phi(a \given d)$ is given directly by the tCNN $\phi$, as shown in Section~\ref{subsec:network} (different options for modeling  $p(c \given a)$ are discussed below).
        Thus, in principle, an estimate of the activity class $\hat{c}$ for a given data segment $d$ is obtained in two steps:
        \begin{enumerate}
            \item[(i)] Compute $\pi$ via the neural network $\phi$: $\pi = \phi(d)$
            \item[(ii)] Compute Eq.~\ref{eq:class-estimate} with these values $\pi$
        \end{enumerate}
        
        In the following, we present two options for modeling $p(c \given a)$: The \emph{Direct Attribute Prediction} (DAP), proposed by \cite{lampert_attribute-based_2014}, and the \emph{Generalized DAP}, that we formally introduce here for the first time, but that was implicitly used by \cite{moya2018_LARHAR,niemann_lara_2020} before in an approximate form. 
        Finally, we describe a novel idea that does not explicitly use the probabilistic model above, but uses a classifier to model the dependency between network outputs $\pi$ and activity classes $c$.

        \paragraph{Direct Attribute Prediction}
            The \emph{Direct Attribute Prediction} (DAP) model \cite{lampert_attribute-based_2014} assumes that each class $c$ has an associated unique attribute representation $a^{(c)}$, i.e.\ there is a deterministic relationship $f(c) = a^{(c)}$ so that 
            $p(a \given c) = \mathds{1}(a = a^{(c)})$.
            By making use of Bayes' theorem, one can rewrite $p(c \given a) = \frac{p(c)}{p(a)} p(a \given c)$. 
            Inserting back into Eq.~\ref{eq:attribute-class-model} gives
            \begin{align}
            p(c \given d) & =  \sum_{a \in A} \frac{p(c)}{p(a)} p(a \given c)\, p_\phi(a \given d)\\
            \intertext{The normalization factor $p(a)$ is constant w.r.t.\ the class. Furthermore, \cite{lampert_attribute-based_2014} propose to use a uniform class prior $p(c)$. Thus, the factor $\frac{p(c)}{p(a)}$ can be ignored for classification, so that the expression simplifies to}
            p(c \given d) & \propto \sum_{a \in A} \mathds{1}(a = a^{(c)})\, p_\phi(a \given d)\\
            & = p_\phi(a^{(c)} \given d),
            \end{align}
            where $a^{(c)} = f(c)$ is the unique attribute representation of class $c$. 
        
        \paragraph{Generalized Direct Attribute Prediction}
            Unfortunately, in HAR, activity classes usually do not have a unique attribute representation. For example, the activity class \emph{take} can involve the attribute \emph{left hand} and/or \emph{right hand}. 
            Thus, the DAP model cannot be used directly.
            
            However, a deterministic relationship in the opposite direction often exists in HAR, i.e., each attribute vector corresponds to exactly one activity class. 
            This mapping $g(a) = c$ can either be defined in advance from prior domain knowledge, or estimated from training data. 
            
            In either way, the distribution becomes $p(c \given a) = \mathds{1}(c = g(a))$, which can be inserted into Eq.~\ref{eq:class-estimate} and \ref{eq:attribute-class-model}. In this case, the maximum-likelihood estimate is computed as 
            \begin{equation}
                \hat{c} = \text{argmax}_c \sum_{\{a \in A  \,|\, g(a) = c \}} p_\phi(a \given d)
                \label{eq:GDAP}
            \end{equation}
             We call this model \emph{Generalized} Direct Attribute Prediction (GDAP).
            
            However, computing the sum in Eq.~\ref{eq:GDAP} exactly is only feasible as long as the number of attribute vectors $|A|$ is small.
            To make classification feasible even when $|A|$ is large,  \cite{moya2018_LARHAR,niemann_lara_2020} propose to compute the sum not exactly, but to compute the maximum-likelihood approximation\footnote{More precisely, \cite{moya2018_LARHAR,niemann_lara_2020} do not compute $\hat{a} = \text{argmax}_a \, p_\phi(a \given d)$, but $\hat{a} = \text{argmin}_a \, ||a - \pi||_2$ after normalizing $a$ and $\pi$, hence we call this model  \emph{nearest neighbor approximation}.}
            $\hat{a} = \text{argmax}_a \, p_\phi(a \given d)$, and return $\hat{c} = g(\hat{a})$. That is, their approximation assumes that the sum in Eq.~\ref{eq:GDAP} is dominated by its largest term, similar to the approximation done in \cite{ramirez2010probabilistic} for goal recognition.

        \paragraph{Classification-based methods}
            The former methods either require to sum over all attribute vectors (which quickly becomes infeasible, as the number of possible attribute vectors grows exponentially with the number $K$ of attributes), or need to make strong assumptions that might induce a large error. 
            
            Therefore, we propose a novel approach to obtain activity classes, shown in Fig.~\ref{fig:method}.
            The key insight is to view the neural network outputs $\pi_1, \dots, \pi_n$ not as parameters of $p_\phi(a \given d)$, but as high-level \emph{features} or \emph{latent representations} that can be used as input  of a classifier that learns the function $c = h(\pi)$.
            
            In this case, an estimate of the activity class $\hat{c}$ is for a given data segment $d$ obtained in two steps: 
           (i) Compute the neural network output $\pi = \phi(d)$; (ii) Compute the activity class estimate $\hat{c} = h(\pi)$.
            Specifically, we use Quadratic Discriminant Analysis (QDA), a Hidden Markov Model with Gaussian observation model (HMM) and Random Forests (RF) as classifiers. 
            
            In addition to the potentially higher computational efficiency during classification, this concept has two key advantages over the probabilistic models introduced before: 
            \begin{itemize}
            \item It allows to account for bias in the neural network, i.e., the shallow classifier can produce a correct classification even when the network output $\pi$ does not assign a high probability to attribute vectors associated with the true class. 
            \item Integrating process step information as an additional feature of the classifier becomes straightforward, and does not require re-training of the complete neural network when additional process information becomes available, as discussed next.
            \end{itemize}

    \subsection{Making use of Process Knowledge}
    \label{subsec:using-process-knowledge}
    
        \begin{figure}
            \centering
            \includegraphics[width=\columnwidth]{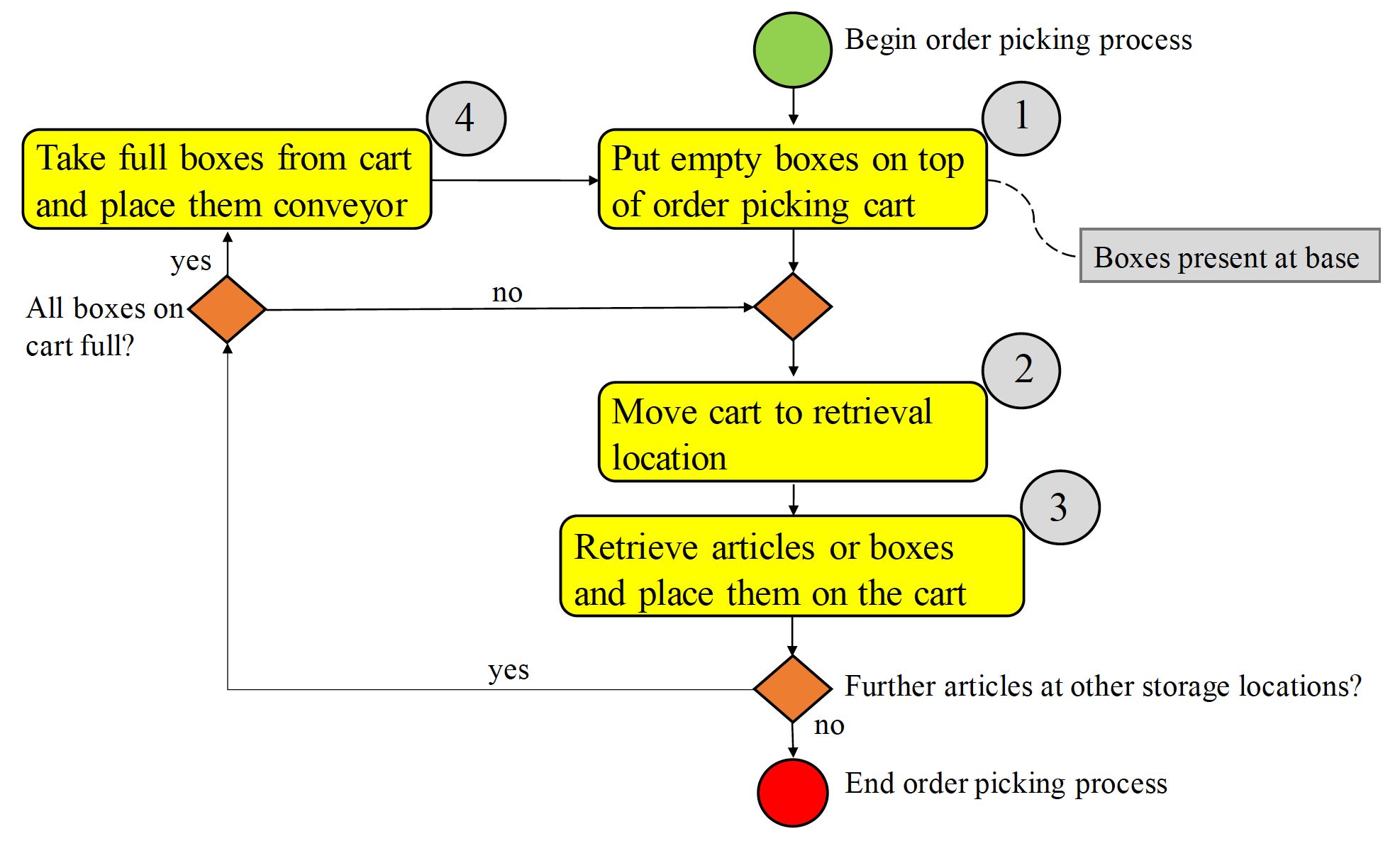}
            \caption{The Business Process Model (BPM) for the logistics scenario.}
            \label{fig:bpm}
        \end{figure}

        In highly structured domains like manual-work processes, the overall activity sequence can often be separated into distinct \emph{process steps}, see Fig.~\ref{fig:attribute_based_representation}. As a concrete example, consider the Business Process Model (BPM) in Figure \ref{fig:bpm} that describes a simple order-picking process in a warehouse. At each point in time, the subject is performing exactly one of these process steps.
        
        Here, we assume that an estimate of the current process step is available. 
        For example, in the warehouse scenario, the worker needs to scan each article before putting it inside a box for further transportation or packing. These scanning events, thus, allow recognizing process-step transitions accurately. 
        In other domains, information about process transitions might not be available directly, but can be estimated by causal models, like Computational State-Space Models \cite{kruger2014computational}.
        
        
        In principle, such process step information could be used directly as additional input to the attribute network $\phi$. However, this would require re-training of the network, which can be computationally very costly. Additionally, the original training data still need to be available to allow re-training, which might not be the case when using a pre-trained network as a black-box model. Furthermore, we assume that the data-attribute dependency that is modeled by the neural network is independent of the process step---how the sensor observations relate to poses and movements of the subject, i.e., the attributes, does not change when the subject is executing different sub-tasks.
        
        \begin{table}[t]
        \centering
        \small
        \begin{tabular}{p{6mm}p{6mm}p{6mm}p{6mm}p{7mm}p{10mm}p{7mm}p{7mm}}
          \toprule
        BPM state & Stand & Walk & Cart & Handle (up) & Handle (center) & Handle (down) & Sync. \\ 
          \midrule
        S0 & 0.266 & 0.375 & 0.000 & 0.047 & 0.195 & 0.031 & 0.086 \\ 
          S1 & 0.000 & 0.053 & 0.000 & 0.342 & 0.474 & 0.132 & 0.000 \\ 
          S2 & 0.011 & 0.000 & 0.757 & 0.006 & 0.215 & 0.011 & 0.000 \\ 
          S3 & 0.061 & 0.000 & 0.000 & 0.239 & 0.534 & 0.166 & 0.000 \\ 
          S4 & 0.000 & 0.095 & 0.000 & 0.167 & 0.500 & 0.238 & 0.000 \\ 
           \bottomrule
        \end{tabular}
        \caption{Distribution of activity classes for each BPM state.}
        \label{tbl:class-priors}
        \end{table}
        
        Instead, we assume that only the relationship between \emph{attributes and activity classes} depends on the process step. 
        Therefore, we propose to include the process step information into the shallow classifier that predicts activity classes $c$ from network outputs $\pi$.
        More specifically, we make use of the process knowledge in the different classifiers in the following ways:
        \begin{itemize}
            \item In the Random Forest, we add the process step as an additional input feature.
            \item For the QDA, we estimate different priors $p_s(c)$ for each process step $s$, but use a single, shared Gaussian likelihood $p(\pi \given c)$, because the class priors differ substantially in each process step (see Table \ref{tbl:class-priors}), but the amount of data is insufficient to estimate the parameters of individual likelihoods for each process step.
            \item For the HMM, we train different transition models $p_s(c_t \given c_{t-1})$ for each process step $s$, to account for the different system dynamics of the different process steps. Similar to the QDA above, the model uses a shared Gaussian observation likelihood $p(\pi \given c)$.
        \end{itemize}

\section{Experimental Evaluation}

    \subsection{Dataset: LARa}
        
        The Logistic Activity Recognition Challenge (LARa) dataset \cite{niemann_lara_2020,Friedrich2020-LARa} is used to evaluate our approach. The dataset recreates three warehousing scenarios in a constrained environment, ensuring natural motion and resemblance to reality \cite{reining2018_TFSANHOPAUMC}.
        This dataset contains measurements of a marker-based MoCap system, called LARa-MoCap, and on-body devices, called LARa-OB, from $14$ humans performing activities in the intra-logistics. LARa provides recordings from the 3D joint-poses and the 3D linear and angular acceleration of subjects performing eight activities. Joint poses are recorded with a rate of $200$ Hz, and linear and angular accelerations with $100$ Hz.  The activities are common activities in the intra-logistics: \textit{Standing}, \textit{Walking}, \textit{Moving Cart}, \textit{Handling (upwards)}, \textit{Handling (centred)}, \textit{Handling (downwards)}, \textit{Synchronization} and \textit{None}. Here, we experimented with LARa-MoCap. For LARa-MoCap, there are $22$ joints. For each joint, LARa provides the $3D$ pose. These are all centered with respect to the lower back of a subject. In general, LARa provides $714$ min of annotated recordings being a large annotated dataset for HAR.
        
        The LARa dataset consists of of non-overlapping training, validation, and testing sets. For LARa-MoCap, the validation and testing sets contain recordings from subjects $[5,11,12]$. The training set contains recordings from the other eight subjects of LARa-MoCap. Our experiments have used the test set, which contains annotations of process steps from one of the three warehouse scenarios.

    \subsection{Results}
    
        \begin{table}[t]
        \centering
        \footnotesize
        \begin{tabular}{lrrrr}
          \toprule
          & \multicolumn{2}{l}{No Process Info} & \multicolumn{2}{l}{With Process Info}\\
          \cmidrule(lr{.75em}){2-3} \cmidrule(lr{.75em}){4-5}
        Classifier & Acc. & F1 & Acc. & F1 \\ 
        \midrule
        GDAP & 0.636 & 0.628 & -- & -- \\ 
        NN & 0.621 & 0.614 & --  & -- \\ 
        QDA & 0.670 & 0.650 & \textbf{0.721} & 0.711 \\ 
        HMM & 0.685 & 0.672 & 0.719 & 0.708 \\
        RF & \textbf{0.688} & \textbf{0.674} & 0.710 & \textbf{0.715} \\ 
        \bottomrule
        \end{tabular}
        \caption{\footnotesize{Experimental results, best results are printed in boldface.
        For each model, we compared a baseline case where no process information is available and the case where the correct BPM state at each time step. 
        For GDAP and NN, information about process states cannot be integrated directly.
        GDAP: Generalized Direct Attribute Prediction; NN: Nearest neighbor-approximation of GDAP; QDA: Quadratic Discriminant Analysis; HMM: Hidden Markov Model; RF: Random Forest.}}
        \label{tbl:results}
        \end{table}
        
        We compared the different methods, introduced in Section \ref{subsec:attributes-to-class}), for obtaining activity classes from the  tCNN$_{attribute}$ output. Specifically, we compared our GDAP model, the nearest-neighbor (NN) approximation of GDAP proposed by \cite{niemann_lara_2020}, and three classification-based methods (QDA, HMM and RF). All experiments were performed in R \cite{rteam}, and we used the \texttt{randomForest} \cite{randomForest} with $500$ trees for fitting RFs.
        
        We used a pre-trained tCNN for attribute prediction for all experiments, with the architecture described in Section \ref{subsec:network}. The pre-trained tCNN can be found in \cite{Friedrich2020-LARa}. The network was trained on the full LARA-MoCap dataset, as described in \cite{niemann_lara_2020}.
        Furthermore, for the classification-based methods, we investigated the case where the correct BPM state (see Fig.~\ref{fig:bpm}) is provided to the models for each time step, see Section \ref{subsec:using-process-knowledge} for how this feature is utilized in the different classifiers.

        Table \ref{tbl:results} shows the HAR performance of the different methods.
        When the currently performed BPM state is not provided to the models, the classification-based methods (QDA, HMM, RF) all outperform the probabilistic models (GDAP, NN).
        Of the classification-based methods, the RF achieves highest accuracy (0.668) and F1 score (0.674) -- higher than prior results by \cite{niemann_lara_2020}  (NN, with 0.621 accuracy and 0.621 F1 score). 
        
        To assess the information contained in the BPM state, we first computed the mutual information $I(C,S) = H(C) - H(C \given S)$  of activity classes $c$ and BPM states $s$ (where $H(C)$ the marginal entropy of activity classes and $H(C \given S)$ is the conditional entropy): On the test data, we obtain $H(S) = 2.44\,\text{bit}$ and $H(C \given S)=1.55\,\text{bit}$. Thus, knowledge of the BPM states provides $I(C,S) = 0.89\,\text{bit}$ of additional information for classification. 
        This finding is reflected in the classification performance: The performance of all models increases when they are provided with the correct BPM state at each timestep (see Table \ref{tbl:results}). 
        Again, all three classifiers show similar performance, with RF having the highest F1 score (0.715), and QDA having the highest accuracy (0.721).

        \begin{figure}
            \centering
            \includegraphics[width=0.85 \columnwidth]{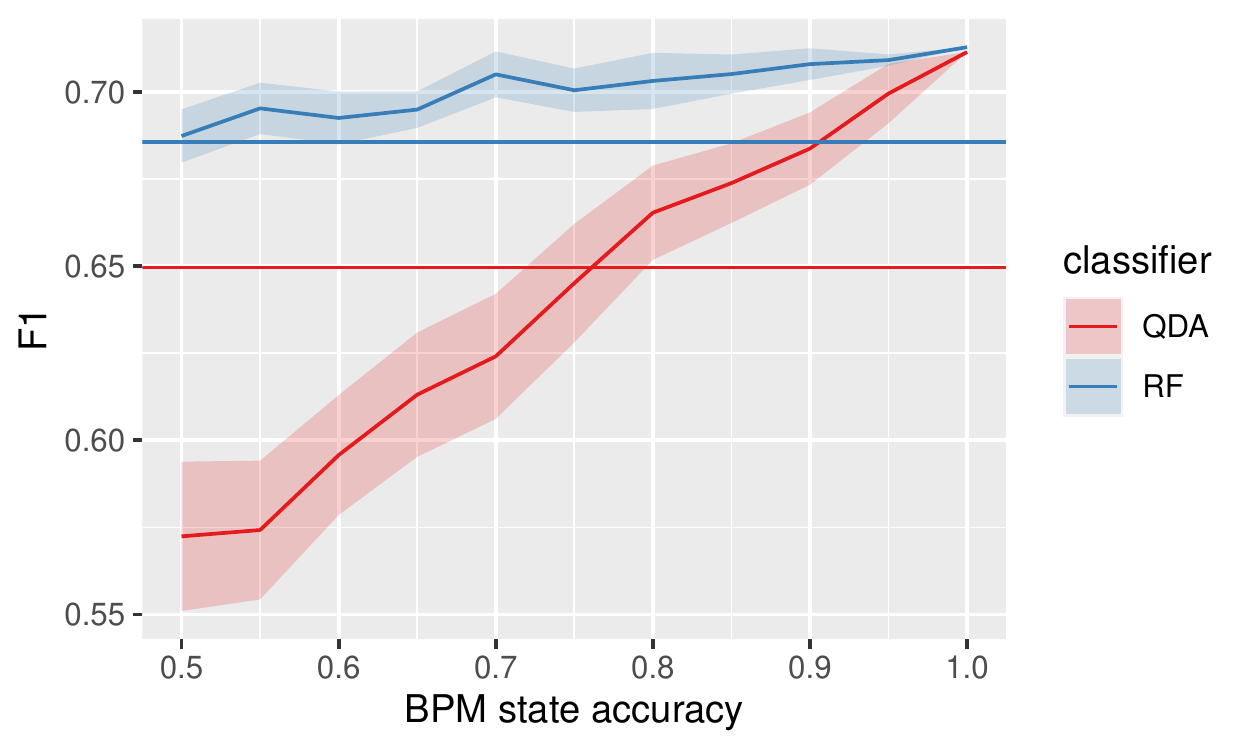}
            \caption{\footnotesize{F1 score w.r.t.\ accuracy of BPM state prediction. Horizontal lines indicate the baseline performance without BPM state information. Each experiment has been repeated 10 times, the ribbons indicate standard deviation of the F1 score. The RF is more robust regarding inaccurate state information than the QDA.}}
            \label{fig:state-accuracy-f1}
        \end{figure}

        Next, we investigated how inaccuracies in the provided BPM state influence HAR performance.
        Specifically, we introduced inaccuracies into the BPM state feature as follows: With probability $p$, each BPM state $s_t$ is replaced by a different state, chosen uniformly among the remaining states $S \setminus s_t$. 
        Thus, the noisy state sequence has an accuracy of $1-p$. We varied the parameter $p$ between $0$, i.e., perfectly accurate state information,  and $0.5$, in steps of $0.05$, and investigated how this affected the F1 score of the HAR models. 
        
        Fig.~\ref{fig:state-accuracy-f1} shows the results for this experiment. For both the QDA and the RF, the F1 score decreases when the BPM state estimate becomes less accurate. However, the RF is more robust in that regard: Even when the accuracy of the BPM state estimate is only 50\%, the RF is not worse than the baseline RF that does not make use of the BPM state at all -- in contrast to the QDA, which requires a state accuracy of at least 80\% to outperform the QDA baseline. 
        
        In summary, the results indicate that providing the shallow classifiers with context information in the form of the currently performed BPM state increases the overall HAR performance, even when the BPM state is not always accurate. 

\section{Discussion and Conclusion}

    In this paper, we proposed a HAR system that combines a deep neural network and a shallow classifier. The system predicts activity classes from the neural network output, i.e., from the approximations of the posterior distribution over attributes. 
    Our empirical evaluation on a HAR task in logistics shows that this combined method can increase HAR performance, compared to previous methods \cite{niemann_lara_2020}.
    Furthermore, this system allows integrating additional context information (like an estimate of the current process step, as used here) directly, without re-training the neural network. 
    We showed that providing an estimate of the current process step as an additional feature can increase HAR performance, even when the process-step estimation is not always correct.
    
    However, these results are preliminary: Our evaluation is based on only a subset of the complete LARa dataset, as process step annotations were only available for one of the three scenarios contained in the dataset. Furthermore, we just simulated the noise on the process step estimate. 
    Future work needs to confirm our findings for a more realistic case where the process step is estimated from external sources, as well as on additional HAR datasets.
    On a technical level, future work will investigate using an additional neural network layer for predicting activity classes from attributes instead of using a shallow classifier, thus allowing end-to-end-training of the complete HAR system.

    Depending on the sensor setup and the specific domain, additional context information, other than the current process step, might be available, e.g., currently handled objects or the location of the subjects. Our proposed system architecture allows using such features directly in a systematic way. Investigating such cases is another topic for future work. 
    
    Finally, the BPM can be seen as a form of prior domain knowledge that we aim to exploit. 
    So far, we only made use of the fact that different BPM states have different prior probabilities of activities, but the BPM also provides information about the \emph{dynamics} of states and action. In the future, we will explore how such information can serve to generate symbolic precondition-effects models of the system dynamics, e.g., Computational State-Space Models \cite{kruger2014computational}.

\bibliographystyle{named}
\bibliography{bibliography}

\end{document}

%% file: figures/attribute_based_representation.tex
\begin{figure}[!t]
    \centering

    \begin{tikzpicture} [x=0.7cm,y=0.7cm]

        \tikzstyle{node1}=[text=black, font=\small \bfseries];
        \tikzstyle{node2}=[text=black, font=\small \bfseries];
        \tikzstyle{node3}=[text=black, font=\tiny];
        \tikzstyle{node4}=[text=black, font=\scriptsize];
        \tikzstyle{node5}=[text=black, font=\scriptsize ];
        \tikzstyle{arrow1} = [->,line width=0.06cm]
        \tikzstyle{circle1}=[circle,draw=black, minimum size=0.1cm, line width=0.2mm, inner sep=0pt]
        \tikzstyle{circle2}=[circle,draw=black, minimum size=0.15cm, line width=0.1mm, inner sep=0pt, fill=yellow]
        \tikzstyle{circle3}=[circle,draw=black, minimum size=0.15cm, line width=0.1mm, inner sep=0pt, fill=white]
        \tikzstyle{rectangle1}=[rectangle, minimum width=2cm, minimum height=1cm, text width=1.7cm, font=\scriptsize]

        \node [node1] at (-1.5,0.5){Steps};
        
        \draw [line width=0.05mm,fill=tucol4, opacity=0.3](3.0,1.0)--(0.0,1.0)--(0.4,0.5)--(0.0,0,0)--(3.0,0.0)--(3.5,0.5)--cycle;
        \node [node1] at (1.8,0.5){...};
        
        \draw [line width=0.05mm,fill=tucol4, opacity=0.3](3.4 + 4.0,1.0)--(-0.4 + 4.0,1.0)--(0.0 + 4.0,0.5)--(-0.4 + 4.0,0.0)--(3.4 + 4.0,0.0)--(3.9 + 4.0,0.5)--cycle;
        \node [node1] at (1.8 + 4.0,0.5){Transporting Box};

        \draw [line width=0.05mm,fill=tucol4, opacity=0.3](3.0 + 7.6,1.0)--(0.0 + 8.0,1.0)--(0.4 + 8.0,0.5)--(0.0 + 8.0,0.0)--(3.0 + 7.6,0.0)--(3.5 + 7.6,0.5)--cycle;
        \node [node1] at (1.8 + 8.0,0.5){...};

        \node [node1] at (-1.1,0.5 - 1.50){Activities};

        \draw [line width=0.05mm,fill=tucol4, opacity=0.3](1.0,1.0 - 1.5)--(0.0,1.0 - 1.5)--(0.4,0.5 - 1.5)--(0.0,0.0 - 1.5)--(1.0,0.0 - 1.5)--(1.4,0.5 - 1.5)--cycle;
        \node [node2] at (0.8,0.5 - 1.50){...};

        \draw [line width=0.05mm,fill=tucol4, opacity=0.3](1.5 + 1.2,1.0 - 1.5)--(0.0 + 1.2,1.0 - 1.5)--(0.4 + 1.2,0.5 - 1.5)--(0.0 + 1.2,0.0 - 1.5)--(1.5 + 1.2,0.0 - 1.5)--(1.9 + 1.2,0.5 - 1.5)--cycle;
        \node [rectangle1] at (1.3 + 1.55,0.5 - 1.5) {Standing};
        

        \draw [line width=0.05mm,fill=tucol4, opacity=0.3](1.5 + 2.9,1.0 - 1.5)--(0.0 + 2.9,1.0 - 1.5)--(0.4 + 2.9,0.5 - 1.5)--(0.0 + 2.9,0.0 - 1.5)--(1.5 + 2.9,0.0 - 1.5)--(1.9 + 2.9,0.5 - 1.5)--cycle;
        \node [rectangle1] at (0.8 + 3.8,0.5 - 1.50) {Walking};

        \draw [line width=0.05mm,fill=tucol4, opacity=0.3](1.5 + 1.2 + 3.4,1.0 - 1.5)--(0.0 + 1.2 + 3.4,1.0 - 1.5)--(0.4 + 1.2 + 3.4,0.5 - 1.5)--(0.0 + 1.2 + 3.4,0.0 - 1.5)--(1.5 + 1.2 + 3.4,0.0 - 1.5)--(1.9 + 1.2 + 3.4,0.5 - 1.5)--cycle;
        \node [rectangle1] at (1.3 + 1.2 + 3.9,0.5 - 1.5) {Cart};
        
        \draw [line width=0.05mm,fill=tucol4, opacity=0.3](1.5 + 2.4 + 3.9,1.0 - 1.5)--(0.0 + 2.4 + 3.9,1.0 - 1.5)--(0.4 + 2.4 + 3.9,0.5 - 1.5)--(0.0 + 2.4 + 3.9,0.0 - 1.5)--(1.5 + 2.4 + 3.9,0.0 - 1.5)--(1.9 + 2.4 + 3.9,0.5 - 1.5)--cycle;
        \node [rectangle1] at (0.8 + 2.4 + 4.7,0.5 - 1.5) {Handling};


        \draw [line width=0.05mm,fill=tucol4, opacity=0.3](1.5 + 1.2 + 6.8,1.0 - 1.5)--(0.0 + 1.2 + 6.8,1.0 - 1.5)--(0.4 + 1.2 + 6.8,0.5 - 1.5)--(0.0 + 1.2 + 6.8,0.0 - 1.5)--(1.5 + 1.2 + 6.8,0.0 - 1.5)--(1.9 + 1.2 + 6.8,0.5 - 1.5)--cycle;
        \node [rectangle1] at (0.8 + 1.2 + 7.7,0.5 - 1.5) {Synchr.};
        
        \draw [line width=0.05mm,fill=tucol4, opacity=0.3](1.0 + 2.4 + 7.3,1.0 - 1.5)--(0.0 + 2.4 + 7.3,1.0 - 1.5)--(0.4 + 2.4 + 7.3,0.5 - 1.5)--(0.0 + 2.4 + 7.3,0.0 - 1.5)--(1.0 + 2.4 + 7.3,0.0 - 1.5)--(1.4 + 2.4 + 7.3,0.5 - 1.5)--cycle;
        \node [node2] at (0.8 + 2.4 + 7.3,0.5 - 1.5){...};



        \node [circle3] (c5) at (0.6 + 1.4,-1.7-0.6) {};
        \node [circle2] (c6) at (0.6 + 1.4,-2.05-0.6) {};
        \node [circle2] (c7) at (0.6 + 1.4,-2.4-0.6) {};
        \node [circle3] (c5) at (0.6 + 1.4,-2.75-0.6) {};
        \node [circle3] (c6) at (0.6 + 1.4,-3.1-0.6) {};
        \node [circle2] (c7) at (0.6 + 1.4,-3.45-0.6) {};

        \node [circle2] (c5) at (0.6 + 3.3,-1.7-0.6) {};
        \node [circle3] (c6) at (0.6 + 3.3,-2.05-0.6) {};
        \node [circle3] (c7) at (0.6 + 3.3,-2.4-0.6) {};
        \node [circle2] (c5) at (0.6 + 3.3,-2.75-0.6) {};
        \node [circle3] (c6) at (0.6 + 3.3,-3.1-0.6) {};
        \node [circle2] (c7) at (0.6 + 3.3,-3.45-0.6) {};

        \node [circle2] (c5) at (0.6 + 5.0,-1.7-0.6) {};
        \node [circle3] (c6) at (0.6 + 5.0,-2.05-0.6) {};
        \node [circle2] (c7) at (0.6 + 5.0,-2.4-0.6) {};
        \node [circle2] (c5) at (0.6 + 5.0,-2.75-0.6) {};
        \node [circle2] (c6) at (0.6 + 5.0,-3.1-0.6) {};
        \node [circle3] (c7) at (0.6 + 5.0,-3.45-0.6) {};

        \node [circle3] (c5) at (0.6 + 6.7,-1.7-0.6) {};
        \node [circle3] (c6) at (0.6 + 6.7,-2.05-0.6) {};
        \node [circle2] (c7) at (0.6 + 6.7,-2.4-0.6) {};
        \node [circle3] (c5) at (0.6 + 6.7,-2.75-0.6) {};
        \node [circle2] (c6) at (0.6 + 6.7,-3.1-0.6) {};
        \node [circle3] (c7) at (0.6 + 6.7,-3.45-0.6) {};

        \node [circle3] (c5) at (0.6 + 8.4,-1.7-0.6) {};
        \node [circle2] (c6) at (0.6 + 8.4,-2.05-0.6) {};
        \node [circle2] (c7) at (0.6 + 8.4,-2.4-0.6) {};
        \node [circle2] (c5) at (0.6 + 8.4,-2.75-0.6) {};
        \node [circle3] (c6) at (0.6 + 8.4,-3.1-0.6) {};
        \node [circle3] (c7) at (0.6 + 8.4,-3.45-0.6) {};

        \node [node1] at (-1.0,-2.3-0.6){Attributes};
        \node [node5] at (10.19,-1.71-0.6){Gait Cycle};
        \node [node5] at (10.17,-2.055-0.6){Torso Rot.};
        \node [node5] at (10.27,-2.44-0.6){Right Hand};
        \node [node5] at (10.202,-2.7-0.6){Left Hand};
        \node [node5] at (10.311,-3.1-0.6){Handy Unit};
        \node [node5] at (10.074,-3.45-0.6){No item};

        \node [node1] at (-1.55,-2.3-2.6-0.6){Data};
        \begin{axis}[
                    axis lines=middle,
                    ymin=0,ymax=10,
                    xshift=0.0cm,yshift=-5.1cm ,
                    axis y line = none,
                    width=9cm,
                    xlabel={Time [s]},
                    xticklabels={,,},
          ]
          \addplot[ samples=150, domain=-pi:4*pi] ({x}, {1+0.2*sin(4*deg(x)) + 0.05*sin(128*deg(x))});
          \addplot[ samples=150, domain=-pi:4*pi] ({x}, {1.5+0.2*sin(8*deg(x))});
          \addplot[ samples=150, domain=-pi:1*pi] ({x}, {2.0+0.2*sin(16*deg(x))});
          \addplot[ samples=150, domain=pi:2*pi] ({x}, {2.0+0.0*sin(16*deg(x))});
          \addplot[ samples=150, domain=2*pi:4*pi] ({x}, {2.0+0.1*sin(32*deg(x))});
          \addplot[ samples=150, domain=-pi:0.1*pi] ({x}, {2.5+0.1*sin(32*deg(x)) +0.1*sin(16*deg(x))});
          \addplot[ samples=150, domain=0.1*pi:1*pi] ({x}, {2.3+0.065*x});
          \addplot[ samples=150, domain=1*pi:3*pi] ({x}, {2.5+0.1*sin(8*deg(x)) +0.1*sin(4*deg(x))});
          \addplot[ samples=150, domain=3*pi:4*pi] ({x}, {2.5-0.05*sin(32*deg(x)) +0.1*sin(8*deg(x))});
        \end{axis}

        \draw [line width=0.05mm,fill=tucol5, opacity=0.15](1.2 + 1.5+3.5,1.0 - 5.1-0.6)--(0.0 + 1.5+3.5,1.0 - 5.1-0.6)--(0.0 + 1.5+3.5,0.0 - 6.3-0.6)--(1.2 + 1.5+3.5,0.0 - 6.3-0.6)--cycle;

        \node (net1) at (0.5 + 1.5+3.6,-4.3-0.6) {};
        \node (net2) at (0.5 + 1.5+3.6,-3.5-0.6) {};
        \draw  [arrow1](net1) edge (net2);

        \draw [line width=0.05mm,fill=green, opacity=0.10](3.6,0.0)--(1.2, -0.5)--(9.5,-0.5)--(7.4,0.0)--cycle;
        
        \node (net1) at (5.6, 0.15) {};
        \node (net2) at (5.6, -0.65) {};
        \draw  [arrow1](net1) edge (net2);
        
        \node (net1) at (5.6, -2.2) {};
        \node (net2) at (5.6, -1.4) {};
        \draw  [arrow1](net1) edge (net2);

    \end{tikzpicture}

    \caption{\footnotesize{Relationship between attributes, process steps and activity classes. First, attributes are estimated from the raw sensor data. Activities are then classified based on these attributes and the current process step.}}
    \label{fig:attribute_based_representation}
\end{figure}

%% file: figures/method_fig.tex
\begin{figure}[t]
    \centering

    \begin{tikzpicture} [x=1.0cm,y=0.9cm]

        \tikzstyle{node1}=[text=black, font=\normalsize \bfseries];
        \tikzstyle{node2}=[text=black, font=\small \bfseries];
        \tikzstyle{node3}=[text=black, font=\small];
        \tikzstyle{node4}=[text=black, font=\footnotesize];
        \tikzstyle{arrow1} = [line width=0.7]
        \tikzstyle{arrow2} = [line width=0.7]
        \tikzstyle{circle1}=[circle,draw=black, minimum size=0.1cm, line width=0.2mm, inner sep=0pt]
        \tikzstyle{circle2}=[circle,draw=black, minimum size=0.05cm, line width=0.1mm, inner sep=0pt, fill=black]

        

        \draw [fill=tucol8, line width=0.05mm, opacity=0.4] (0.0 + 0.9 ,0.0 + 1.2) rectangle +(1.3,0.8);
        
        \draw [arrow2] (0.0 - 0.3,0.0 + 1.5) edge (0.0 + 0.9,0.0 + 1.5);
        
        \node (label) at (0.0 + 2.0, 0.0 + 2.4)[draw=white, line width=0.0]{
                \includegraphics[width= 0.47\textwidth]{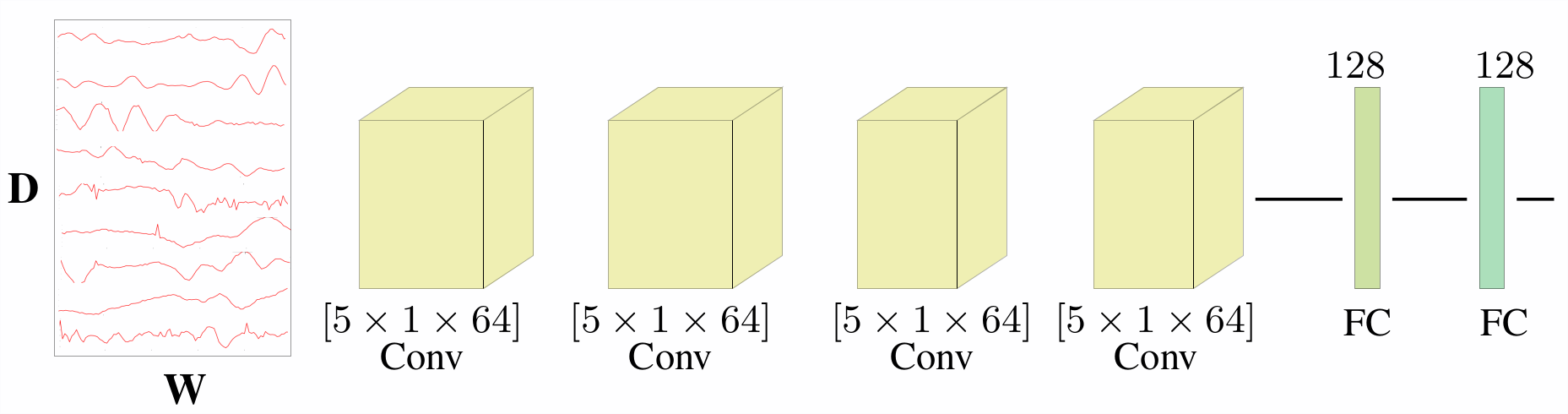}
              };
        \node [node4] at (0.0 - 1.5, 0.0 + 3.75){Sensor Data};
        \node [node4] at (0.0 + 1.5,0.0+3.4){tCNN};
        
        \draw [arrow2] (0.0 + 1.5,0.0 + 1.3) edge (0.0 + 1.5,0.0 + 1.0);
        \draw [arrow2] (0.0 + 6.4,0.0 + 1.3) edge (0.0 + 1.5,0.0 + 1.3);
        \draw [arrow2] (0.0 + 6.4,0.0 + 2.46) edge (0.0 + 6.4,0.0 + 1.3);
        \draw [fill=tucol5, line width=0.05mm, opacity=0.4] (0.0 - 2.9+3.4,0.0 + 0.5) rectangle +(2.0,0.4);
        \node [node3] at (0.0 - 1.5+3.0,0.0 + 0.69){Sigmoid};
        
        \draw [fill=tucol8, line width=0.05mm, opacity=0.4] (0.0 - 2.9+3.0,0.0 - 3.8+3.0) rectangle +(2.8,0.4);
        \draw [arrow2] (0.0 - 1.5+3.0,0.0 - 4.2+4.6) edge (0.0 - 1.5+3.0,0.0 - 4.5+4.6);
        \node [node4] at (0.0 - 1.5+3.0,0.0 - 3.2+3.0){Approximations of $p_\phi(a \given d)$};
        \node [node4] at (0.0 - 1.5+3.0,0.0 - 3.6+3.0){$\pi_1, \pi_2, ..., \pi_K$};
        
        \draw [fill=tucol8, line width=0.05mm, opacity=0.4] (0.0 + 0.6+3.3,0.0 - 3.8+3.0) rectangle +(1.8,0.4);
        \draw [arrow1] (0.0 + 1.5+3.3,0.0 - 3.8+3.0) edge (0.0 + 1.5+3.3,0.0 - 4.3+3.0);
        \draw [arrow1] (0.0 - 1.5+3.0,0.0 - 4.3+3.0) edge (0.0 + 1.5+3.3,0.0 - 4.3+3.0);
        \node [node4] at (0.0 + 1.5+3.3,0.0 - 3.2+3.0){Context Information};
        \node [node4] at (0.0 + 1.5+3.3,0.0 - 3.6+3.0){$S$};
        
        \draw [fill=tucol8, line width=0.05mm, opacity=0.4] (0.0 - 2.9+3.0, 0.0 - 5.3+3.0) rectangle +(2.8,0.8);
        \draw [arrow2] (0.0 - 1.5+3.0,0.0 - 3.8+3.0) edge (0.0 - 1.5+3.0,0.0 - 4.5+3.0);
        \node [node4] at (0.0 - 1.5+3.0, 0.0 - 4.7+3.0){Shallow Classifiers};
        \node [node4] at (0.0 - 1.5+3.0, 0.0 - 5.1+3.0){(QDA, HMM, RF)};
        
        \draw [fill=tucol8, line width=0.05mm, opacity=0.4] (0.0 - 2.1+3.0, 0.0 - 6.8+3.0) rectangle +(1.3, 0.5);
        \draw [arrow2] (0.0 - 1.5+3.0,0.0 - 5.3+3.0) edge (0.0 - 1.5+3.0,0.0 - 5.9+3.0);
        \node [node4] at (0.0 - 1.5+3.0, 0.0 - 6.1+3.0){Activity Class Estimate};
        \node [node4] at (0.0 - 1.5+3.0, 0.0 - 6.55+3.0){$\hat{C}$};
            
    \end{tikzpicture}

    \caption{\footnotesize{Proposed hybrid activity recognition architecture. The neural network provides probabilities $\pi_i$ that each of the attributes $a_i$ is present. The probability vector $\pi$ and the context information (e.g.\ the current process step) is used as input of a shallow classifier that estimates an activity class $\hat{c}$.}}
    \label{fig:method}
\end{figure}